\documentclass{aastex63}
\usepackage{threeparttable}

\shorttitle{Double-Peaked SLSNe}
\shortauthors{Liu et al.}

\graphicspath{{./}{figures/}}

\begin{document}

\title{Magnetar-Driven Shock Breakout Revisited and Implications  for Double-Peaked Type I Superluminous Supernovae}

\correspondingauthor{Liang-Duan Liu; He Gao}
\email{liuliangduan@bnu.edu.cn, gaohe@bnu.edu.cn}

\author[0000-0002-8708-0597]{Liang-Duan Liu}
\affiliation{Department of Astronomy, Beijing Normal University, Beijing 100875, China}

\author[0000-0002-3100-6558]{He Gao}
\affiliation{Department of Astronomy, Beijing Normal University, Beijing 100875, China}

\author[0000-0002-7334-2357]{Xiao-Feng Wang}
\affiliation{Physics Department and Tsinghua Center for Astrophysics, Tsinghua University, Beijing 100084, China}
\affiliation{Beijing Planetarium, Beijing Academy of Sciences and Technology, Beijing 100044, China}

\author[0000-0002-2898-6532]{Sheng Yang}
\affiliation{Department of Astronomy, The Oskar Klein Center, Stockholm University, AlbaNova, 10691 Stockholm, Sweden}

\begin{abstract}

The discovery of early bumps in some type-I superluminous supernovae (SLSNe-I) before the main peaks offers an important clue to their energy source mechanisms. In this paper, we updated an analytic magnetar-powered model for fitting the multi-band light curves of double-peaked SLSNe-I: the early bump is powered by magnetar-driven shock breakout thermal emission, and the main peak is powered by a radiative diffusion through the SN ejecta as in the standard magnetar-powered model.  Generally, the diffusive luminosity is greater than the shock breakout luminosity at the early time, which makes the shock breakout bumps usually not clearly seen as observed. To obtain a clear double-peaked light curve, inefficient magnetar heating at early times is required. This model is applied to three well-observed double-peaked SLSNe-I (i.e., SN2006oz, LSQ14bdq, and DES14Xtaz).  We find that a relative massive SN ejecta with $M_{\mathrm{ej}} \simeq 10.2-18.1 M_{\odot}$ and  relative large kinetic energy of SN ejecta $E_{\mathrm{sn}} \simeq (3.8-6.5) \times 10^{51}$ erg are required, and the thermalization efficiency of the magnetar heating is suppressed before $t_{\mathrm{delay}}$, which are in the range of $\simeq 15- 43$ days.
The model can well reproduce the observed light curves, with a reasonable and similar set of physical parameters for both the early bump and the main peak, strengthening support for magnetar-powered model. In the future, modeling of the double-peaked SLSNe-I will become more feasible as more events are discovered before the early bump.


\end{abstract}


\section{Introduction} \label{sec:intro}

High cadence, unbiased wide-field optical transient surveys have uncovered a new class of highly luminous transients, so-called superluminous supernovae (SLSNe; \citep[see][for a review]{Gal19}). Based on the absence or presence of hydrogen spectral features during the maximum light, SLSNe are usually classified into hydrogen-poor (SLSNe-I) or hydrogen-rich (SLSNe-II) subclasses, respectively. The high luminosity of SLSNe-II are mainly thought to be powered by the interaction of the SN ejecta with a dense  hydrogen-rich circumstellar medium (CSM) surrounding the progenitor \citep{Smi07,Che11}. However, the power mechanisms of SLSNe-I have not yet reached a consensus \citep[see][for reviews]{Mor18,Wang19}.  

Two popular scenarios have been proposed to account for the huge luminosities of SLSNe-I. In the first scenario, the SN ejecta is surrounded by a massive  hydrogen-poor CSM. Interaction between SN ejecta and CSM would convert kinetic energy of the SN ejecta into thermal radiation \citep{Che11,Gin12,Cha12,Cha13}. Whereas the second scenario invokes energy input from a central engine, such as spin-down of a newborn rapidly rotating magnetar \citep{Kas10,Woo10} or fallback accretion into a central compact object \citep{Dex13,Mor18b,Gao20, Lin20}. Many previous works have shown that both scenarios could interpret the general shape of the photometric light curves of SLSNe-I, i.e, matching the peak time, peak luminosity, rising slope and the decaying slope\citep[e.g.,][]{Cha13,Ins13,Wang15,Pra17,Yu17,Liu17,Nic17,Mor18b}.

Additional observation details are required to discriminate energy source models of SLSNe-I. The light curves of some SLSNe-I exhibit a double-peaked structure on the rise, i.e., with a precursor ``bump” prior to the primary one \citep{Lel12,Nic15,Smi16}.  Such an early-time excess emission has not been reported for any SLSN-II so far. \cite{Nic16} did a systematic search for the early bump in the light curves of all published SLSNe before that time. They found that 8 of 14 objects  with relevant early data are found to have plausible early bumps, suggesting that these bumps may be common in SLSNe-I. Moreover, the recent sample from the Dark Energy Survey shows that early bumps are not ubiquitous in the SLSNe-I \citep{Ang19}. The discovery of such double-peaked light curves in some SLSNe-I could  offer new insights into the energy source mechanisms of SLSNe. 

The precise nature of the early bumps remains unclear. The early bump is too narrow and bright to be powered by the radioactive decay of $^{56}$Ni. Under the framework of CSM interaction scenario, some works attempt to explain the early bump by greatly increasing the model complexity, e.g. by introducing multiple collisions \citep{Liu18b}, by invoking ionization change in massive CSM \citep{Mor12}, or by invoking a huge explosion energy ($\sim 10^{52}$ erg) and a large progenitor radii (a few hundred $R_\odot$) to power the early bump via post-shock cooling of extended stellar material \citep{Piro15,Nic15,Piro20}. On the other hand, by contrast, \cite{Kas16} suggested that the double-peaked behavior of SLSNe-I is naturally expected within the magnetar scenario, where the early bump is the emission from the breakout of the magnetar wind-driven shock wave, while the main peak of light curve is powered by diffusive radiation from direct magnetar heating. 

The goal of this paper is to explore whether the magnetar-powered model can satisfactorily fit the real SLSNe-I double-peaked light curves, and whether the required magnetar and ejecta parameters are feasible. For this purpose, building on the work of \cite{Kas16}, we introduce a characteristic timescale $t_{\mathrm{delay}}$ to quantitatively describe the delay time due to inefficient thermalization of the magnetar wind. This allows us to provide a set of more reader-friendly expressions for magnetar wind-driven shock breakout, which is more convenient to compare with the observation data in a large region of parameter space. 

The paper is organized as follows: the well-observed sample of SLSNe-I with double peaks are described in Section \ref{sec:sample}. We describe in detail how a fast rotating magnetar produces a double-peaked light curve of SLSNe in Section \ref{sec:model}. The results of the light curve fitting are presented in Section \ref{sec:Fitting}. Finally, we give the discussions and conclusions of this paper in Section \ref{sec:Discussion}. Numerical quantities are denoted $Q_x = Q/10^x$ is cgs units, unless otherwise explicitly stated.

\section{Double-peaked sample of SLSNe-I} \label{sec:sample}

In order to explore the nature of double-peaked SLSNe-I, sufficient observational data of the early bump are required. We searched the literature for double-peaked SLSNe-I with several epochs of photometry catching the early bump. Eventually, three SLSNe-I are selected into our sample, and their names and main properties are collected in Table \ref{Tab:sample}. Specifically, \cite{Lel12} identified a 10 days-long bump in SLSN-I SN2006oz with multi-band photometry, followed by a 25-30 days-rise to the main peak. \cite{Nic15} reported a similar bump in LSQ14bdq, which has a better time resolution but only $r$-band photometry is available for the early bump. The difference in the $r$-band peak magnitude between the early bump and main peak is $ M_{\mathrm{bump}}^{\mathrm{max}}-M_{\mathrm{peak}}^{\mathrm{max}} \sim 2$ mag. Another well-observed case is DES14X3taz. \cite{Smi16} reported the pre-peak bump for DES14X3taz, which was detected simultaneously in the DES $griz$ bands 20 days prior to the rise of the main peak. The peak luminosity of the early bump reaches $\sim 30 \% $ that of the main peak. \cite{Smi16} found that the temperature of DES14X3taz rapidly cooled from 22,000 to 8000 K during the early bump (i.e., within about 2 weeks). Note that \cite{Lun18} reported two possible double-peaked SLSNe-I (i.e., PS1-11aib and PS1-13) in the Pan-STARRS1 SLSNe sample. Moreover, \cite{Ang19} recently found several double-peaked candidates from DES SLSNe sample. However, all of these events do not show clearly distinct bumps like that seen in the above three sample (SN 2006oz, LSQ14bdq, and DES14x3taz), thus we did not include them into our dataset. All data used in our fits are publicly available from the Open Supernova Catalog \footnote{\url{https://sne.space/}} \citep[OSC;][]{Gui17}.  


\begin{table}[ht]
    \centering
    \caption{List of Double-Peaked SLSNe-I in Our Sample}
    \begin{threeparttable}
    \begin{tabular}{lcccccl}
    \hline \hline
      Object   &  Redshift & Filters\tnote{a} & $M^{\mathrm{max}}_{r,\mathrm{bump}}$ & $M^{\mathrm{max}}_{r,\mathrm{peak}}$  & E(B-V) \tnote{b} & Reference \\  \hline 
      SN2006oz   & 0.376 & $ugri$ & $-19.32$  & $-21.41$  & 0.0403 & \cite{Lel12} \\
      LSQ14bdq   & 0.345 & $r$ & $-20.06$ & $-21.89$  & 0.0559 & \cite{Nic15} \\
      DES14X3taz   & 0.608 & $griz$ & $-19.38$  & $-21.78$  & 0.022 & \cite{Smi16} \\ 
      \hline
    \end{tabular}
      \begin{tablenotes}
    \item[a] This column lists the filters have the early bump detection of each SLSN in our sample.
    \item[b] This column lists the host galaxy reddening.
  \end{tablenotes}
    \label{Tab:sample}
\end{threeparttable}
\end{table}

\section{Model} \label{sec:model}

In this section, we describe how a fast rotating magnetar produces a double-peaked light curve of SLSNe in detail. The magnetar wind could affect the dynamics of the SN ejecta, creating a high-pressure bubble that drives shock heating \citep{Kas16,Suz17,Suz20}. If the spin energy of the magnetar is large enough, the shock could become radiative near the surface of SN ejecta and produce an early-time bump. The main peak of light curve in the magnetar model is produced by the diffusion of radiation from the magnetar heating.

For a core-collapse SN (CCSN) with the initial kinetic energy of the SN ejecta  $E_{\rm sn}$, ejecta mass $M_{\rm ej}$, from progenitor of initial radius $R_{\mathrm {ej,i}}$. The characteristic  velocity of the SN ejecta could be estimated as 
\begin{equation}
    v_{\rm sn} = \left( \frac{2 E_{\rm sn}}{M_{\rm ej}} \right)^{1/2} \simeq 3.2 \times 10^{3} \mathrm{\ km \ s}^{-1} E_{\rm sn,51}^{1/2} M_{\rm ej,10}^{-1/2},
\end{equation}
where  $M_{\rm ej,10}=M_{\rm ej}/10M_{\odot}$. The expansion timescale is
\begin{equation}
    t_{\mathrm{ex}} = \frac{R_\mathrm{ej,i}}{v_{\mathrm{sn}}} \simeq 3 \times 10^{4} \mathrm{\ s \ } R_\mathrm{ej,i,13} E_{\rm sn,51}^{-1/2} M_{\rm ej,10}^{1/2}.
\end{equation}
where $R_{\mathrm {ej,i,13}} $ is the initial radius in unit of $10^{13}$ cm.

After a certain time of expansion, the SN ejecta would enter into the homologous expansion phase when $r=v t$. Based on the numerical simulations, the  density profiles of SN ejecta are expected to have an inner, flat region and an outer steep power-law region as follows (\citep{Che89, Mat99}):
\begin{equation} \label{Eq:rho_ej}
\rho _{\text{ej}}\left( v,t\right) =\left\{
\begin{array}{lc}
\zeta _{\rho }\frac{M_{\text{ej}}}{v_{\text{tr}}^{3}t^{3}}\left( \frac{r}{v_{%
\text{tr}}t}\right) ^{-\delta }, & v<v_{\text{tr}}, \\
\zeta _{\rho }\frac{M_{\text{ej}}}{v_{\text{tr}}^{3}t^{3}}\left( \frac{r}{v_{%
\text{tr}}t}\right) ^{-n}, & v \geq v_{\text{tr}},%
\end{array}%
\right.
\end{equation}%
where $\delta <3$ and $n>5$ are required to ensure that the mass and energy do not diverge. The typical values for CCSNe are $\delta=1, n=10$ \citep{Che89}. The SN ejecta density is continuously across the transition point between the inner and outer profiles, thus the transition velocity is 
\begin{equation}
   v_{\text{tr}}=\zeta _{v}\left( \frac{ E_{\text{SN}}}{M_{\text{ej}}} \right)^{1/2} .
\end{equation}

The numerical coefficients $\zeta_\rho$ and $\zeta_v$ depend on the values of $\delta$ and $n$, which are given in Eqs (6) and (7) of \cite{Kas16}. For $\delta=1, n=10$ one has  $\zeta_\rho=0.124$ and   $\zeta_v=1.69$.
 
For a  magnetar that is born in a CCSN, the
bubble of the magnetic field and relativistic particles  produced by
the proto-magnetar expands in the freely expanding SN ejecta.  The initial rotational energy of the magnetar is $E_{\mathrm{m}}=(1/2)I\Omega^2 \simeq 2 \times 10^{52} P_{\mathrm{ms}}^{-2}$ erg, where $\Omega= 2 \pi/P$ is the angular velocity, $P_{\mathrm{ms}}$ is the magnetar initial period in units of ms, and a magnetar moment of inertia $I$, of $10^{45}$ g cm$^2$ has been adopted. If the magnetic dipole radiation is the dominant mechanism for magnetar spin-down, then the spin-down luminosity can be given by 
\begin{equation} \label{Eq:Ld}
    L_{\mathrm{sd}} = \frac{E_{\mathrm{m}}}{t_{\mathrm{sd}}}\frac{1}{(1+t/t_{\mathrm{sd}})^2} ,
\end{equation}
where $t_{\mathrm{sd}}$ is the spin-down timescale. The high pressure from central energy injection from the magnetar creates a cavity and sweeps the SN ejecta into a thin shell. The central overpressure produced by the energy deposition blows a bubble in the SN ejecta, which is similar to the dynamics investigated in the context of pulsar wind nebulae \citep[e.g,][]{Che92,Che05}. In order to calculate the dynamical evolution of magnetar driven wind, we use the basic conversation equations  of mass, momentum, and energy of the shock dynamics as
\begin{equation}
    \frac{ d M_{\mathrm{sh}}}{d t} = 4 \pi R_{\mathrm{sh}}^2 \rho_{\mathrm{ej}} 
    (v_{\mathrm{sh}}-v_{\mathrm{ej}}),
\end{equation}
\begin{equation}
     M_{\mathrm{sh}} \frac{d v_{\mathrm{sh}}}{d t } = 4 \pi R_{\mathrm{sh}}^2 \left[p_{\mathrm{b}} -\rho_{\mathrm{ej}} 
    (v_{\mathrm{sh}}-v_{\mathrm{ej}})^2 \right],
\end{equation}
\begin{equation}
    \frac{d ( 4 \pi R_{\mathrm{sh}}^3 p_{\mathrm{b}})}{ d t} = L_{\mathrm{sd}}(t) -p_{\mathrm{b}}4 \pi R_{\mathrm{sh}}^2 \frac{d R_{\mathrm{sh}}}{ dt },
\end{equation}
where $p_{\mathrm{b}}$ is the pressure in the magnetar-driven bubble,  $M_{\mathrm{sh}}$ is the mass of shell. Owing to that the gas is assumed to be radiation pressure dominated, the adiabatic index $\hat{\gamma}=4/3$. The shock velocity $v_{\mathrm{sh}}=d R_{\mathrm{sh}}/ dt$  and  the local ejecta expansion velocity $v_{\mathrm{ej}}= R_{\mathrm{sh}}/t$. The density profile of the SN ejecta $ \rho_{\mathrm{ej}} $ was taken from equation (\ref{Eq:rho_ej}). Given the initial conditions, one can solve $R_{\mathrm{sh}}$, $ M_{\mathrm{sh}}$, and $ p_{\mathrm{b}}$ numerically.

Initially, the injected energy for the magnetar can be assumed at a constant rate, $L_{\mathrm{sd,i}}=E_{\mathrm{m}}/t_{\mathrm{sd}}$, the
radius of the swept-up shell has an analytic self-similar power-law solution as \citep{Che05}
\begin{equation} \label{Eq:Rsh1}
    R_{\mathrm{sh}}= v_{\mathrm{tr}}  t_{\mathrm{sh,tr}} \left( \frac{t}{t_{\mathrm{sh,tr}}} \right)^{\alpha},   \text{ \ for  \ } t<t_{\mathrm{sh,tr}},
\end{equation}
where the temporal index $\alpha=(6-\delta)/(5-\delta)$, and the time takes the shock to move through the inner region then reach the transition velocity is \citep{Che05,Kas16}
\begin{equation}
    t_{\mathrm{sh,tr}} = \zeta_{\mathrm{tr}} \left( \frac{E_{\mathrm{sn}}}{E_{\mathrm{m}}} \right) t_{\mathrm{sd}},
\end{equation}
where $ \zeta_{\mathrm{tr}} $ is a coefficient, which is given in Eq(14) of \cite{Kas16}. For $\delta=1, n=10$ one has  $\alpha=5/4$ and   $\zeta_{\mathrm{tr}}=2.19$. One may note that the temporal index $\alpha>1$,  which means if the energy input continues, the expansion will be accelerated. The velocity of the shock is 
\begin{equation}
    v_{\mathrm{sh}}=\frac{d R_{\mathrm{sh}}}{ d t}= \alpha  v_{\mathrm{tr}}  \left( \frac{t}{t_{\mathrm{sh,tr}}} \right)^{\alpha -1} ,
\end{equation}
and at radius $R_{\mathrm{sh}}$ the local ejecta expansion velocity $v_{\mathrm{ej}}= R_{\mathrm{sh}}/t$. When the shock propagates though inner ejecta, the shock moves $25 \%$ faster the local ejecta velocity.

The above self-similar dynamical evolution only holds for the shock remained in the inner ejecta, i.e., $t<t_{\mathrm{sh,tr}}$.  As long as the energy input continues, the expansion will accelerate, and the shell mass increases.  The condition that the bubble  can reach the transition velocity is  $t_{\mathrm{sd}} \gtrsim t_{\mathrm{sh,tr}}$, which is equivalent to 
\begin{equation}
    E_{\mathrm{m}} \gtrsim 2.2 E_{\mathrm{sn}}.
\end{equation}
If the initial magnetar rotational energy is about twice greater than the kinetic energy in the SN, the shock is able to expand through the transition point in the SN  density profile. For times $t>t_{\mathrm{sh,tr}}$, the shock propagates into the steep outer ejecta. The dynamical evolution is determined by the acceleration of a shell of fixed mass, the radius of the shock can be expressed as \citep{Ost71,Che05} 
\begin{equation}  \label{Eq:Rsh2}
    R_{\mathrm{sh}}= v_{\mathrm{tr}}  t_{\mathrm{sh,tr}} \left( \frac{t}{t_{\mathrm{sh,tr}}} \right)^{3/2},   \text{ \ for  \ } t_{\mathrm{sh,tr}} \leq t < t_{\mathrm{sd}}.
\end{equation}

For the sake of simplicity, we assume that the energy input ceases when $t>t_{\mathrm{sd}}$, then the bubble will no longer overtake the materials. The decreases in pressure of the bubble can be roughly assumed as adiabatical and thus the shock radius is regarded as nearly free expansion, so that 
\begin{equation} \label{Eq:Rsh3}
    R_{\mathrm{sh}}= R_{\mathrm{sh,sd}} \left( \frac{t}{t_{\mathrm{sd}}} \right),   \text{ \ for  \ }   t_{\mathrm{sd}}<t,
\end{equation}
where $R_{\mathrm{sh,sd}}= v_{\mathrm{tr}}  t_{\mathrm{sh,tr}} (t_{\mathrm{sd}}/t_{\mathrm{sh,tr}})^{3/2}$ is the size of shock at $t_{\mathrm{sd}}$. 

Given the dynamical evolution of magnetar-driven shock, we estimate the local heating rate by \citep{Kas16,Li16}
\begin{equation}
    H_{\mathrm{sh}}(t)=\frac{1}{2} \frac{d M_{\mathrm{sh}}}{d t} (v_{\mathrm{sh}}-v_{\mathrm{ej}})^2.
\end{equation}
Substituting the density profile of ejecta and the solutions of the  dynamical evolution, the shock heating rate can be rewritten as 
\begin{equation}
    H_{\mathrm{sh}}(t)=\left\{
\begin{array}{ll}
H_{\mathrm{sh,tr}}, & t<t_{\mathrm{sh,tr}}, \\
H_{\mathrm{sh,tr}} \left( \frac{t}{t_{\mathrm{sh,tr}}} \right)^{\frac{3-n}{2}}, & t_{\mathrm{sh,tr}} \leq t < t_{\mathrm{sd}} \\
0, & t_{\mathrm{sd}}\leq t,
\end{array}%
\right.
\end{equation}
where 
\begin{equation}
    H_{\mathrm{sh,tr}}=\frac{2\pi \zeta_\rho}{(5-\delta)^3}\frac{M_{\mathrm{ej}} v_{\mathrm{tr}}^2}{t_{\mathrm{sh,tr}}} \sim 3.7 \times 10^{43} \mathrm{ erg \ s}^{-1} E_{\mathrm{m,51}} t_{\mathrm{sd,5}}^{-1},
\end{equation}
$H_{\mathrm{sh,tr}}$ represents the local heating rate when the shock remains in the inner flat region, and $t_{\mathrm{sd,5}}=t_{\mathrm{sd}}/5$ days. The local heating falls off with time as $H_{\mathrm{sh}} \propto t^{(3-n)/2}$, because the shock enters the outer steep region where the pre-shock density significantly decreases with radius.

Initially, un-shocked SN ejecta stratified above the spherical shell is dense enough to trap the thermal photons. The magnetar-driven shock will become radiative when the shock moves to the region with an optical depth $\tau \sim c/v_{\mathrm{sh}}$ \citep{Kas16}. In the outer ejecta, the optical depth from radius $r$ to the surface can be expressed  
\begin{equation}
    \tau (r,t) =\frac{c}{v_{\mathrm{tr}}} \left( \frac{r}{v_{\mathrm{tr} }t} \right)^{-n+1} \left( \frac{t}{t_{\mathrm{diff,tr}}} \right)^{-2},
\end{equation}
and the effective diffusion time is 
\begin{equation}
    t_{\mathrm{diff,tr}} = \left[ \frac{\kappa \zeta_\rho M_{\mathrm{ej}}}{(n-1) v_{\mathrm{tr}}c} \right]^{1/2} ,
\end{equation}
where $1/(n-1)$ is a geometric factor associated with the outer ejecta density profile, and  $\kappa$ is the opacity. Due to the hot temperature during the SN explosion, we adopt a constant scattering opacity. 

The radius for the shock being radiative can be written as 
\begin{equation}
    R_{\mathrm{bo}}(t) = v_{\mathrm{tr}}t_{\mathrm{diff,tr}} \left( \frac{t}{t_{\mathrm{diff,tr}}} \right)^{\frac{n-4}{n-2}},  \text{ \ for \ } t<t_{\mathrm{diff,tr}},
\end{equation}

 If the magnetar-driven shock moves into the steep outer ejecta before $t_{\mathrm{diff,tr}}$, the shock breakout occurs in the steep outer ejecta. The shock speed moves $ \sim 50 \% $ faster than the local ejecta velocity  and this will produce a luminous breakout emission. Otherwise, shock breakout will occur in the flat inner ejecta, the shock speed moves only $\sim 25 \%$ faster than the local ejecta velocity, which will result in a weak emission. Different regimes of magnetar-driven shock emergence, one can see Figure 2 in \cite{Kas16}. 

Equaling the breakout radius $R_{\mathrm{bo}}$ to  the shock radius  $R_{\mathrm{sh}}$, we can determine the time of shock breakout as 
\begin{equation}
    t_{\mathrm{bo}} = t_{\mathrm{sh,tr}} \left( \frac{t_{\mathrm{diff,tr}}}{t_{\mathrm{sh,tr}}} \right)^{1/3} \sim 14.6 \text{  days \ }  E_{\rm sn,51}^{7/12} M_{\rm ej,10}^{1/4} E_{\mathrm{m,51}}^{-2/3} t_{\mathrm{sd,5}}^{2/3} \kappa_{0.2}^{1/6},
\end{equation}
where $ \kappa_{0.2}=\kappa/0.2$ g cm$^{-2}$. In the above equation, we have plugged the fiducial values of parameters (i.g., $\delta=1.0, n=10, \alpha=3/2$). This breakout time $t_{\mathrm{bo}}$ offers an appropriate timescale when the photons diffusion time from the shock-swept region equals to the elapsed time.

Treating the diffusion of photons throughout the SN ejecta under the
one-zone approximation, we use the integral formalism  derived by \cite{Arn82} to calculate the bolometric light curve of magnetar-driven shock heating, which accounts for the early bump as 
\begin{equation}
    L_{\mathrm{bump}}(t) = e^{-(t/t_{\mathrm{bo}})^2} \int_{0}^{t} 2 H_{\mathrm{sh}}(t')\frac{t'}{t_{\mathrm{bo}}^2}  e^{(t'/t_{\mathrm{bo}})^2} d t'
\end{equation}

In order to power a luminous main peak light curve, the magnetar spin-down energy must be thermalized in the SN ejecta. Initially, the magnetar wind is expected to carry away the rotational energy in the form of Poynting flux. Some dissipation processes in the wind would convert the magnetic energy into the particles (e.g.,$e^{\pm}$ pairs) kinetic energy \citep{Bro16}. These accelerated particles will radiate by  synchrotron and inverse Compton emission, producing high-energy photons in X-rays and gamma-rays. If the SN ejecta is optically thick for these high-energy photons, they  would be absorbed by the ejecta offering a heating source.  If the ejecta becomes optically thin at the high-energy bands, the non-thermal high-energy photons may escape as X-rays and gamma-rays \citep{Kot13,Met14}. 

The efficiency of the thermalization process is still poorly understood.  However, the uncertain details of magnetar wind thermalization are critical in the shape of the early time light curve. The non-thermal high-energy photons must interact with the ejecta several times before thermalizing. The dominant thermalization process for these high-energy photons is pair production by $\gamma-\gamma$ interactions \citep{Kot13}. The high scattering optical depth of the ejecta traps these high-energy photons and delays their thermalization processes. 
\cite{Kas16} found that the  breakout emission can not be distinguished from the main peak of the light curve unless there is incomplete thermalization in the shocked wind bubble at the early time (see Figure 7 in \cite{Kas16}). Here we introduce an efficiency factor of the thermalization as 
\begin{equation}
    \xi_{\mathrm{th}}(t)=\left\{
\begin{array}{ll}
0, & t<t_{\mathrm{delay}}, \\
1, & t_{\mathrm{delay}} \leq t, \\
\end{array}%
\right.
\end{equation}
where $t_{\mathrm{delay}}$ is the characteristic timescale to quantitatively describe the delay time of thermalization due to pair production by $\gamma-\gamma$ interactions.
The above equation represents that the thermalization is completely inefficient when $t<t_{\mathrm{delay}}$, while the efficiency is 100 $ \% $ after that. Such kind of inefficient thermalization has little effect on the dynamics of the shock, but will significant reduce the early diffusive luminosity from magnetar heating.

 The bolometric luminosity of the main peak can be written as \citep{Arn82,Cha12,Kas16}
\begin{equation}
    L_{\mathrm{peak}}(t) = e^{-(t/t_{\mathrm{sn}})^2} \int_{0}^{t} 2 \xi_{\mathrm{th}} L_{\mathrm{sd}}(t')\frac{t'}{t_{\mathrm{sn}}^2}  e^{(t'/t_{\mathrm{sn}})^2} \left[ 1- e^{-(t'/t_{\gamma, \mathrm{diff}})^{-2}} \right] d t'
\end{equation}
where timescale $t_{\mathrm{sn}}=(3\kappa M_{\mathrm{ej}}/4 \pi c v_{\mathrm{sn,f}} )^{1/2}$.  Taking the energy injection from the magnetar, the final value of the typical SN ejecta is $ v_{\mathrm{sn,f}}=[2(E_{\mathrm{sn}}+E_{\mathrm{m}})/M_{\mathrm{ej}}]^{1/2}$. The factor $ \left[ 1- e^{-(t/t_{\gamma, \mathrm{diff}})^{-2}} \right]$ accounts for the gamma-ray leakage \citep{Wang15,Dai16}, and  $t_{\gamma, \mathrm{diff}}=(3 \kappa_\gamma M_{\mathrm{ej}}/4 \pi v_{\mathrm{sn,f}}^2)^{1/2}$ is the effective diffusion time for gamma-ray, where $\kappa_{\gamma}$ is the gamma-ray  opacity of the SN ejecta. A large value of $t_{\gamma, \mathrm{diff}}$ means that most of the gamma-ray photons are trapped inside the SN ejecta. 

The two one-zone model light curves were summed to give the composite SN light curve as 
\begin{equation}
    L_{\mathrm{tot}}(t) = L_{\mathrm{bump}}(t) +  L_{\mathrm{peak}}(t)
\end{equation}

In order to model the multi-band observational light curves, we need  to determine the evolution of the temperature and the photosphere.  The evolution of the photosphere depends on the competition between the expansion and the recession in the co-moving coordinate of the ejecta \citep{Liu18}. Initially, the photospheric velocity $v_{\mathrm{ph}}$ decreased with time as the outer layers of the ejecta became transparent. Setting the photosphere to be in the optical depth where $\tau=2/3$, the photospheric radius evolves with time as
\begin{equation}
    R_{\mathrm{ph}}(t) = v_{\mathrm{tr}}t_{\mathrm{sh,ph}} \left( \frac{t}{t_{\mathrm{sh,ph}}} \right)^{\frac{n-3}{n-1}},  \text{ \ for \ } t<t_{\mathrm{sh,ph}},
\end{equation}
where 
\begin{equation}
    t_{\mathrm{sh,ph}}= t_{\mathrm{ph,tr}} \left( \frac{t_{\mathrm{sh,tr}}}{t_{\mathrm{sd}}} \right)^{\frac{n-1}{4}} ,
\end{equation}
$t_{\mathrm{sh,ph}}$ represents the time when the photospheric velocity $v_{\mathrm{ph}}$  receded to the shock velocity $v_{\mathrm{sh}}$, and  $t_{\mathrm{ph,tr}} = [3 \kappa \zeta_{\rho} M_{\mathrm{ej}}/2(n-1)v_{\mathrm{tr}}^2 ]^{1/2} $ denotes the time when the outer ejecta becomes transparent. At time $t>t_{\mathrm{sh,ph}}$, the photosphere locked in the shock radius as $R_{\mathrm{ph}}\propto t$. When the total optical depth in the  SN ejecta reaches $\tau = 2/3$, the SN ejecta would enter the so called ``nebular phase".

Here  we simply assume that the spectrum of the breakout emission and magnetar heating can be approximated by a quasi-blackbody, thus the effective temperature is estimated as 
\begin{equation}
    T_{\mathrm{eff}} = \left( \frac{L_{\mathrm{tot}}}{4 \pi R_{\mathrm{ph}}^2 \sigma_{\mathrm{SB}}} \right)^{1/4},
\end{equation}
where $\sigma_{\mathrm{SB}}$ is the Stefan-Boltzmann constant.

\section{Light-curves fitting} \label{sec:Fitting}

We now apply our model to fit the light curves of three well-observed double-peaked SLSNe.
We use a Markov chain Monte Carlo  fitter  \texttt{emcee}  package \citep{For13} to fit the multi-band light curves. For each light-curve fitting,  we run the code in parallel using 12 nodes and 50,000 iterations, and the first 5,000 iterations are used to burn in the ensemble.

\begin{table}[ht]
    \centering

    \caption{The Fitting Parameters and Priors Used in Our Calculations}
    \begin{tabular}{lccc}
    \hline \hline
     Parameter   &  Prior  & Min  & Max \\ \hline
     $M_{\mathrm{ej}}/M_{\odot}$   & Log-Flat & 0.1  & 100 \\
     $E_{\mathrm{sn}}/10^{51}$ erg & Flat & 0.1 & 10  \\
     $E_{\mathrm{m}}/10^{51}$ erg & Flat & 10 & 30  \\
     $t_{\mathrm{sd}}/ $days & Log-Flat & 1 & 100  \\
     $t_{\mathrm{delay}}/ $days & Flat & 0 & 50  \\
     $t_{\mathrm{shift}} / $ days & Flat & 0 & 50  \\ \hline
   
    \end{tabular}

    \label{Tab:parameter}
\end{table}

The  priors and allowed ranges of the fitting parameters used in our calculations are given in Table \ref{Tab:parameter}.  There are six free parameters in our calculations, including two SN explosion parameters: the ejecta mass $M_{\mathrm{ej}}$ and the kinetic energy of SN ejecta $E_{\mathrm{sn}}$, and two magnetar parameters:  the spin energy of the magnetar $E_{\mathrm{m}}$ and the characteristic timescale of spin-down $t_{\mathrm{sd}}$.  The delay timescale for the inefficiently thermalized of the magnetar heating $t_{\mathrm{delay}}$,  and the last parameter $t_{\mathrm{shift}}$ is the time for the first detection in optical bands relative to the explosion time.

\begin{table}[ht]
    \centering
    \caption{The Medians and $1\sigma$ Bounds of Our Fitting Parameters}
    \begin{tabular}{lcccccc}
    
    \hline \hline
    Object &    $M_{\mathrm{ej}}$    &  $E_{\mathrm{sn}}$ & $E_{\mathrm{m}}$ &   $t_{\mathrm{sd}}$ &  $t_{\mathrm{delay}}$ & $t_{\mathrm{shift}}$\\
         &  $M_{\odot}$ & $10^{51}$ erg &  $10^{51}$ erg & days  & days & days  \\ \hline

    SN2006oz  & $10.24^{+5.91}_{-2.68}$  & $3.80^{+1.35}_{-0.91}$ & $10.92^{+7.48}_{-3.14}$  &  $10.68^{+3.70}_{-2.72}$    & $14.96^{+2.48}_{-1.90}$  & $7.76^{+2.48}_{-1.77}$  \\

     LSQ14bdq  & $18.14^{+1.11}_{-0.90}$  & $5.48^{+0.32}_{-0.24}$ & $11.29^{+0.88}_{-0.76}$  &  $26.54^{+3.41}_{-2.98}$    & $42.32^{+4.15}_{-3.37}$  & $24.12^{+4.45}_{-3.59}$  \\
     
    DES14X3taz  & $13.31^{+0.14}_{-0.14}$  & $6.52^{+0.13}_{-0.13}$ & $10.12^{+0.04}_{-0.02}$  &  $8.98^{+0.24}_{-0.23}$     & $27.02^{+0.24}_{-0.24}$  & $9.72^{+0.24}_{-0.23}$  \\
     \hline
      
    \end{tabular}

    \label{Tab:result}
\end{table} 

\begin{figure}
    \centering
    \includegraphics[width=0.49\textwidth,angle=0]{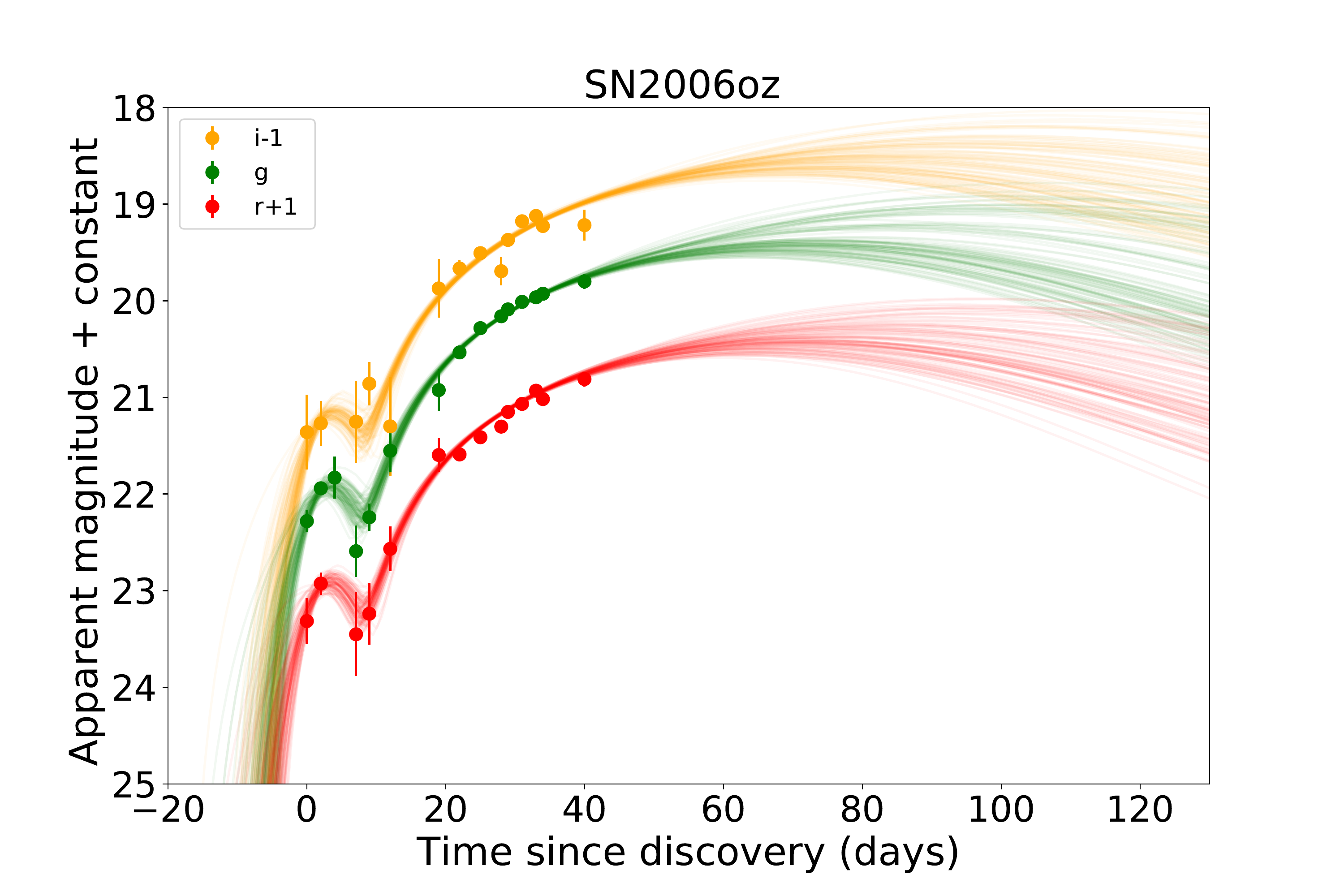}
    \includegraphics[width=0.49\textwidth,angle=0]{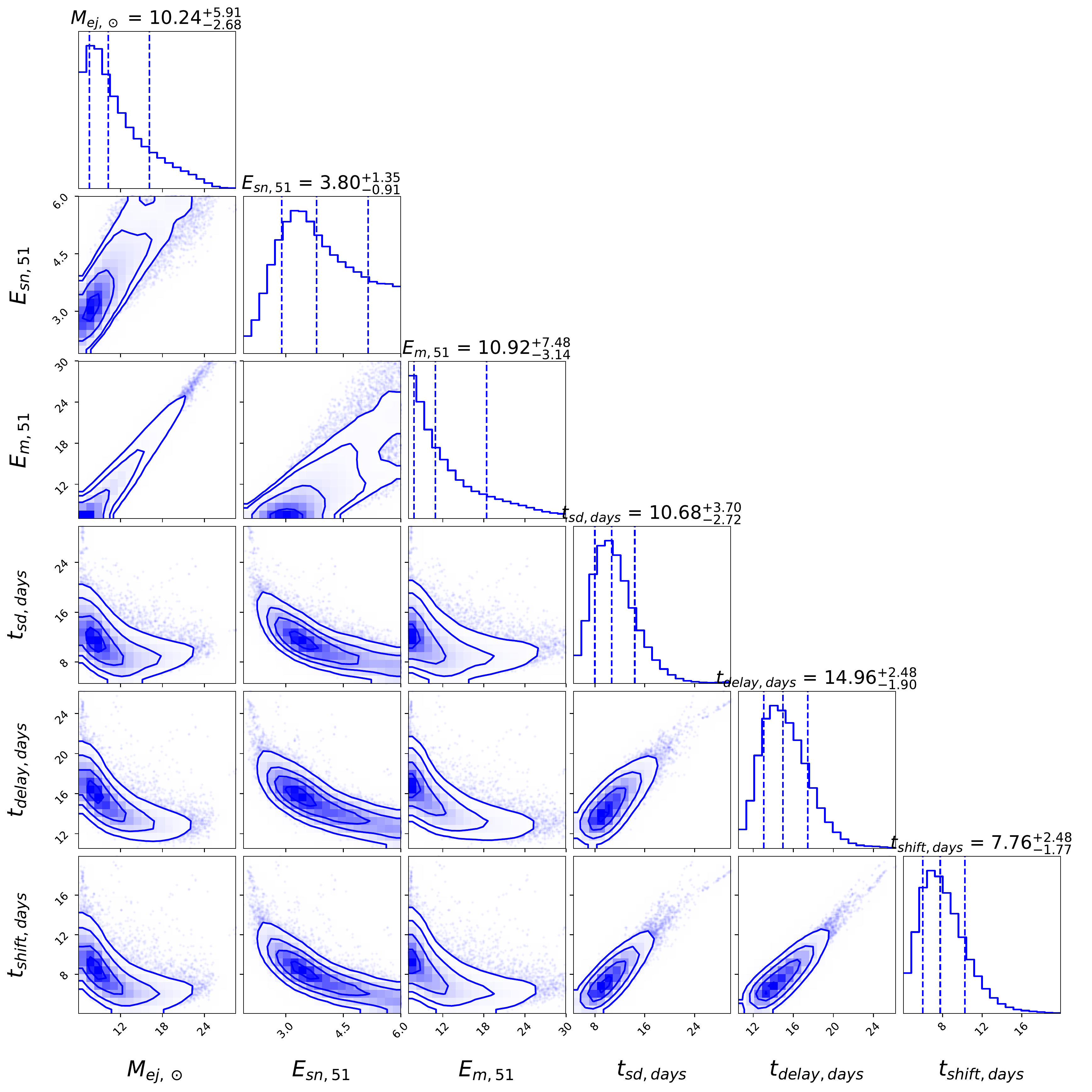}

    \caption{The left panel is fitting result of  magnetar-powered model for the multi-band double-peaked light curves of SN2006oz.  The right panel is the posteriors of the fitting parameters. Medians and 1$\sigma$ ranges are shown. }
    \label{fig:SN2006oz}
\end{figure}

\begin{figure}
    \centering
    \includegraphics[width=0.49\textwidth,angle=0]{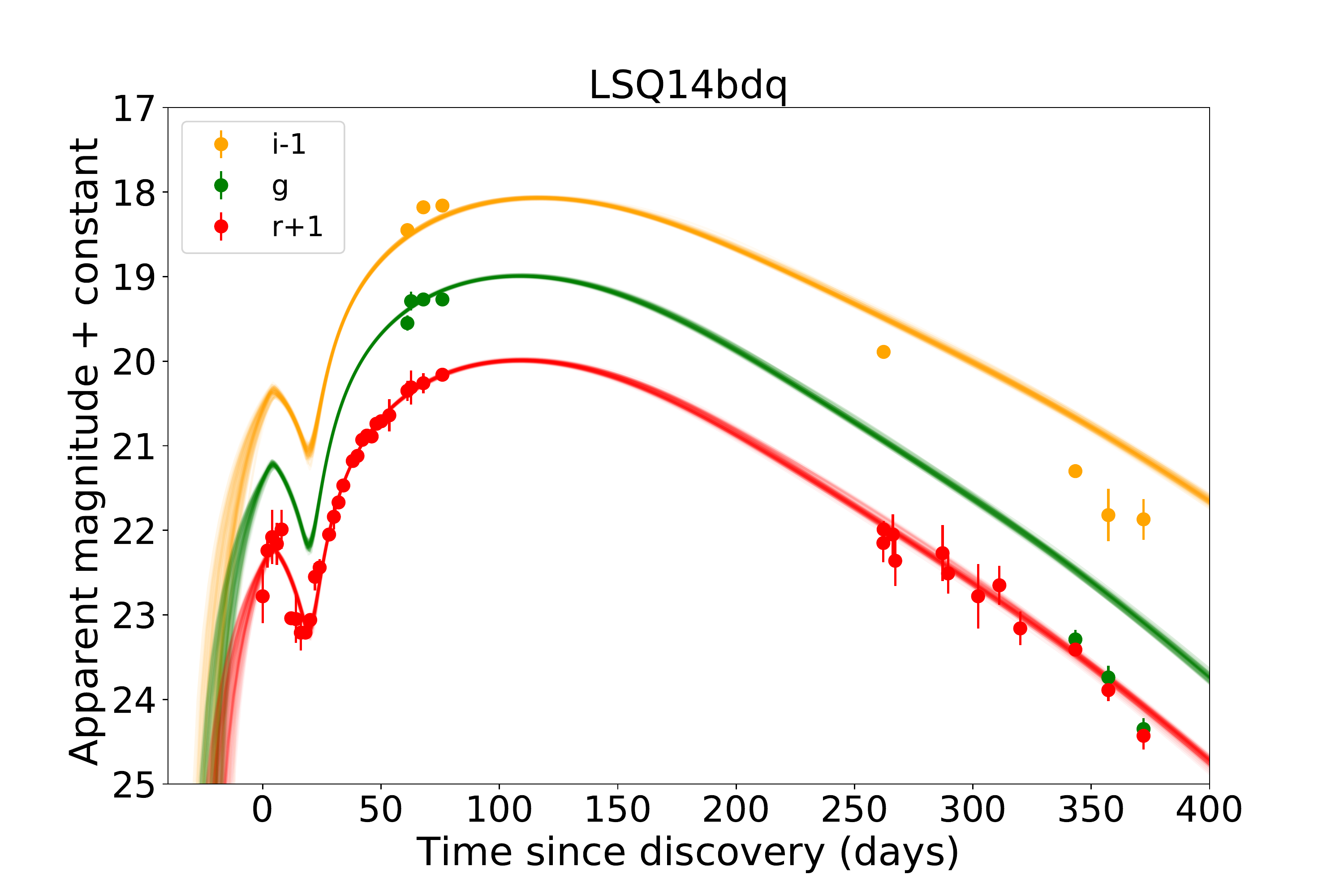}
    \includegraphics[width=0.49\textwidth,angle=0]{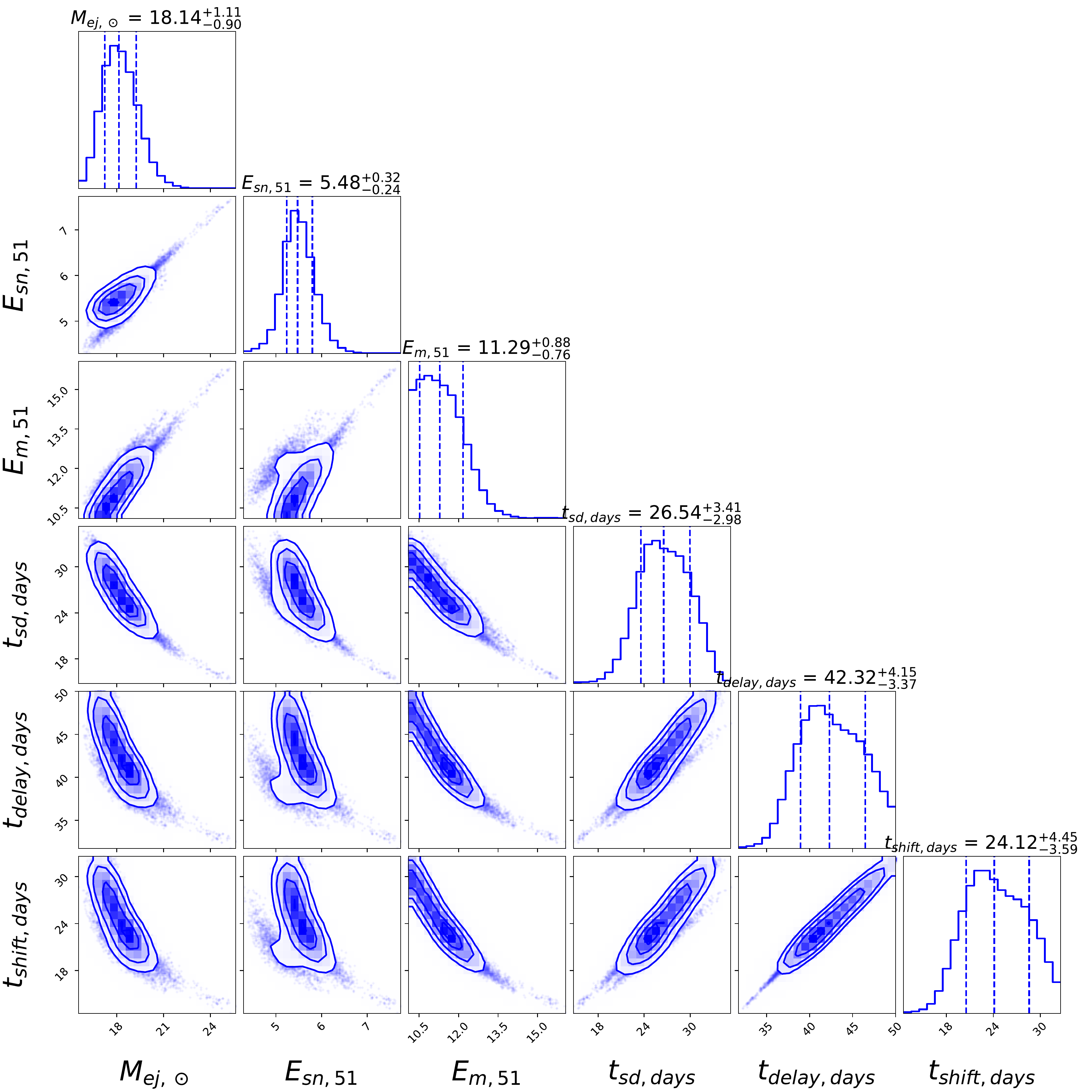}

    \caption{Descriptions of panels are the same as in Figure \ref{fig:SN2006oz}, but for LSQ14bdq.}
    \label{fig:LSQ14bdq}
\end{figure}

\begin{figure}
    \centering
    \includegraphics[width=0.49\textwidth,angle=0]{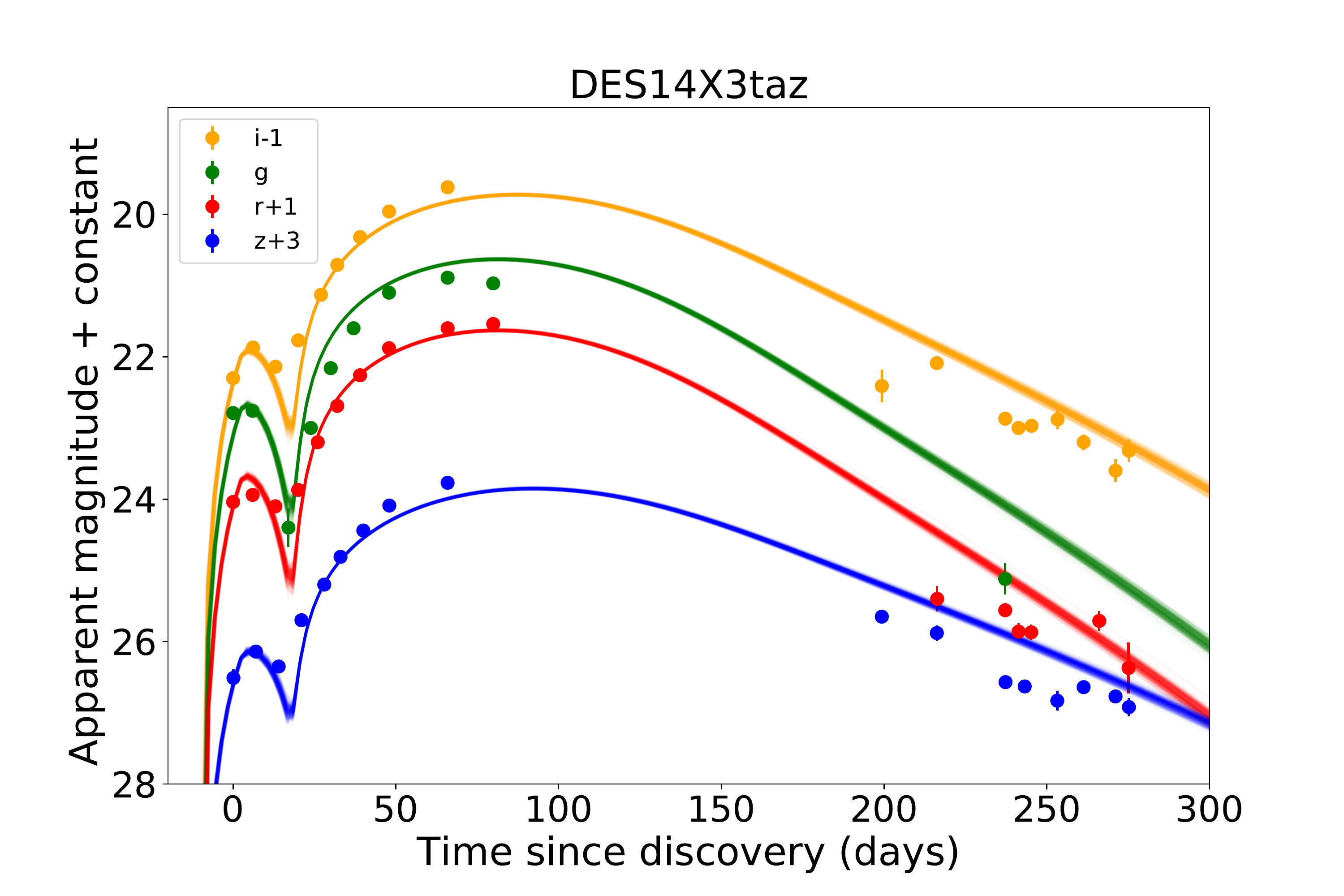}
    \includegraphics[width=0.49\textwidth,angle=0]{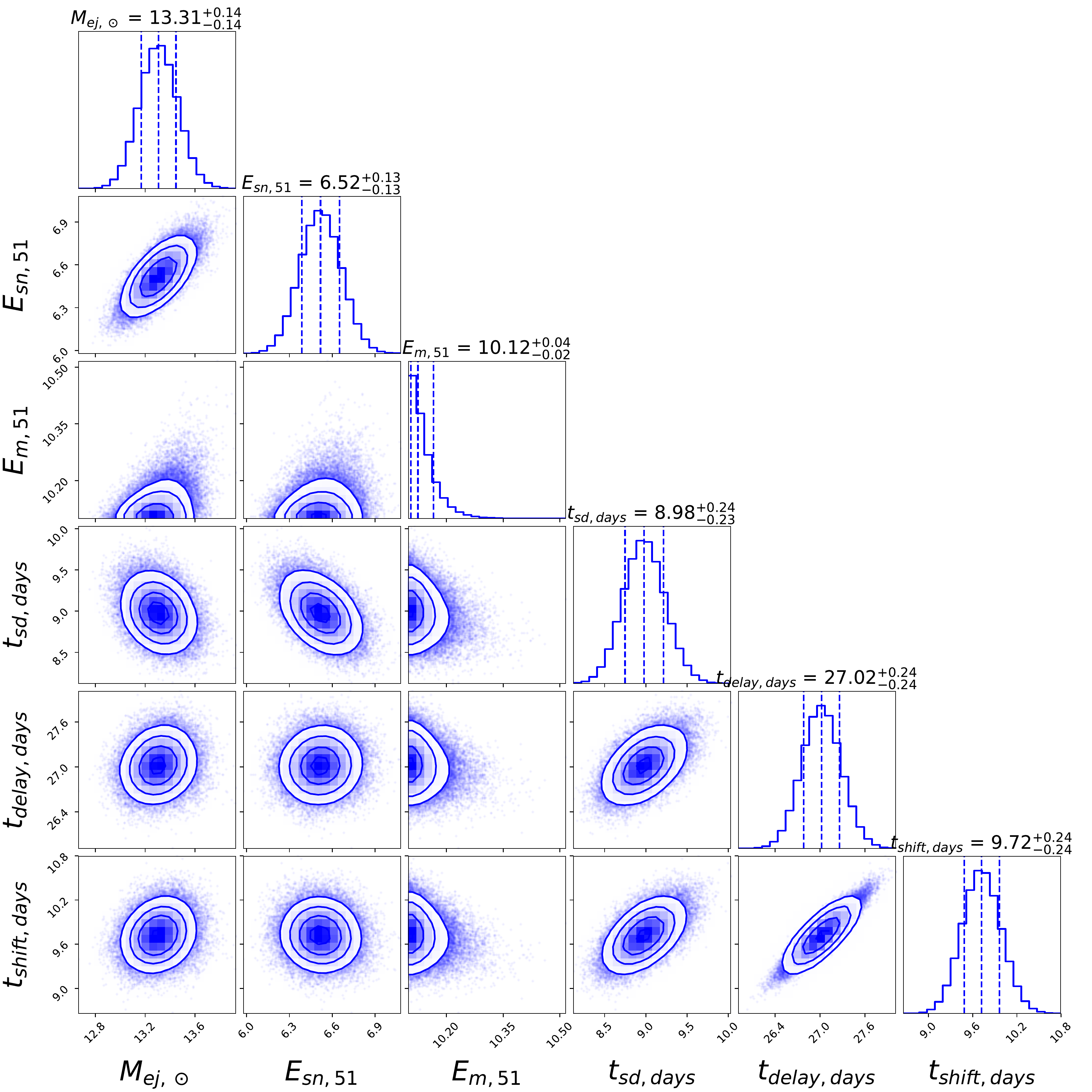}

    \caption{Descriptions of panels are the same as in Figure \ref{fig:SN2006oz}, but for DES14X3taz.}
    \label{fig:DES14X3taz}
\end{figure}

Figure \ref{fig:SN2006oz}-\ref{fig:DES14X3taz} shows the fitting results, the light curves produced by the magnetar-powered model  matched well with the observational data of SN 2006oz, LSQ14bdq, and DES14X3taz. Table \ref{Tab:result} summarizes the parameters and their $1\sigma$
deviations constrained by our fitting. From the corner plots of the  posteriors for the fitting parameters, we find that the parameters can be well constrained. \cite{Nic17} have analyzed the main peaks of the multi-band light curves of these three SLSNe (but they neglected the early bump), by using  the standard magnetar-powered model (without considering the shock breakout emission) in \texttt{MOSFiT}.

As shown in Figure \ref{fig:SN2006oz}, for SN 2006oz, $M_{\mathrm{ej}} =10.2M_{\odot}$,  $E_{\mathrm{sn}}=3.8 \times 10^{51}$ erg,  $t_{\mathrm{sd}} \simeq 10.7$ days. Thus, one can infer the magnetic field as $B \simeq 8.9 \times 10^{13}$ G and the initial period as $P \simeq 1.34$ ms. We discuss our results in comparison with the parameters inferred from \cite{Nic17}  as $M_{\mathrm{ej}}\simeq2.97M_{\odot}$,  $E_{\mathrm{sn}}=2.7 \times 10^{51}$ erg,  $B \simeq 3.2 \times 10^{13}$G, and $P\simeq2.7$ ms. 

LSQ14bdq has a best time resolution of the early bump among three SLSNe. As shown in Figure \ref{fig:LSQ14bdq}, we find that $M_{\mathrm{ej}} =18.2M_{\odot}$, $E_{\mathrm{sn}}=5.49 \times 10^{51}$ erg,  $B \simeq 5.6 \times 10^{13}$ G, and $P\simeq 1.3$ ms are required. 
\cite{Nic17} obtained $M_{\mathrm{ej}}\simeq33.7M_{\odot}$,  $E_{\mathrm{sn}}=25.1 \times 10^{51}$ erg,  $B \simeq 4.9 \times 10^{13}$G, and $P\simeq0.98$ ms.

DES14X3taz has the best multi-band photometric data in our sample, the parameters constrains are the most restrictive see Figure \ref{fig:DES14X3taz}. We find that  $M_{\mathrm{ej}} =13.3M_{\odot}$, $E_{\mathrm{sn}}=6.5 \times 10^{51}$ erg, and $E_{\mathrm{m}} \simeq 1.1 \times 10^{52} $ erg, $B \simeq 1.0 \times 10^{14}$ G, and $P\simeq 1.4$ ms. \cite{Nic17} obtained
 $M_{\mathrm{ej}}\simeq5.41M_{\odot}$,  $E_{\mathrm{sn}}=5.9 \times 10^{51}$ erg,  $B \simeq 3.9 \times 10^{13}$G, and $P\simeq2.4$ ms.


Although the parameters obtained by our model and \cite{Nic17} are consistent to some extent, there are differences in the values of the parameters obtained by different assumptions of magnetar-powered model.
Comparison of our magnetar-powered model with the double-peaked SLSNe SN2006oz, LSQ14bdq, and DES14Xtaz  indicate that these three SNe  had a similar properties of the newly born magnetar (i.e., $B \simeq (0.32-1.0) \times 10^{14}$ G and  $\simeq (1.3- 1.4)$ ms). The ratio between the injected energy and the initial kinetic energy of SN ejecta $E_{\mathrm{m}}/E_{\mathrm{sn}} \simeq 1.6-2.9$. It means that only a sufficiently powerful energy injection can produce double-peaked light curves. The thermalization efficiency of the magnetar heating is suppressed before $t_{\mathrm{delay}}$, which is in the range of $\simeq 15- 43$ days.

\section{Discussion and Conclusions} \label{sec:Discussion}

In this paper, we develop an analytic model involving magnetar-driven shock breakout emission to interpret the double-peaked light curves of SLSNe-I. The main features of magnetar-driven shock breakout emission mechanism are summarized as follows. The peak luminosity of early bump is proportional to the characteristic spin-down luminosity as $L_{\mathrm{bump,peak}} \propto E_{\mathrm{m}} t_{\mathrm{sd}}^{-1}$.  A more massive ejecta mass and a higher explosion energy lead to a longer duration of shock breakout emission. The total heating energy from the shock breakout emission accounts to only a few percent of $E_{\mathrm{m}}$. In general, the diffusive luminosity is greater than the shock breakout luminosity at the early time,which makes the shock breakout bumps usually not clearly double-peaked shape light curves  seen as observed  \footnote{ iPTF13dcc shows early excess emission \citep{Vre17}, but the brightness of the early bump is similar to that of the main peak, which may infer a different nature. \cite{Liu18b} showed that the light curves of iPTF13dcc could be explained with a CSM interaction model with multiple shells.}. To have a clear double-peaked light curve, the thermalization in the shocked wind bubble should be suppressed at early times.

We have presented a set of  Markov chain Monte Carlo model fits to the multi-band light curve of three well-observed double-peaked SLSNe-I and found that the model can well reproduce both the early bump and main peak with the same set of parameters. These results strengthen support for the magnetar-powered model as the energy source of SLSNe. The magnetar-powered model can explain many observed features of several peculiar SNe, for example, \cite{Gre15} presented an ultra-long gamma-ray burst GRB 111209A is associated with a luminous SN 2011kl, both of which are suggested powered by a magnetar; \cite{Dai16} proposed that a rapidly rotating strange quark star accounts for the most luminous SN ASASSN15lh; \cite{Arc17} and \cite{Woo18} mentioned that the spin-down energy of magnetar is one possible long-lasting energy source of iPTF14hls. 

Simultaneous fitting to the whole light curves within the limited model parameters yield that the magnetars have initial periods as  $\simeq (1.3-1.4)$ ms and magnetic field as $B \simeq (0.32-1.0) \times 10^{14}$ G. In order to interpret the double-peaked light curves as observed, a relative massive SN ejecta with $M_{\mathrm{ej}} \simeq 10.2-18.2 M_{\odot}$ and  relative large kinetic energy of the ejecta $E_{\mathrm{sn}} \simeq (3.8-6.5) \times 10^{51}$ erg are required for the SLSNe-I explosions.

The multidimensional numerical simulations of neutrino-driven SNe find
that the upper limit of the kinetic energy provided by neutrinos is $\sim 2 \times 10^{51}$ erg \citep{Ugl12,Ert16,Suk16}. It is smaller than the inferred kinetic energy $(E_{\mathrm{sn}} \gtrsim 3 \times 10^{51}$ erg) of double-peaked SLSNe. This implies  that the SN explosion itself by a different process than the delayed neutrino mechanism, may be exploded by jets \citep{SG17}.  In addition, the inferred spin period of the magnetar is very short $\sim 1.4$ ms, the formation of such rapidly spin magnetar needs the pre-collapse core of the progenitor to rotate at a high rate. \cite{Soker16,Soker17}  argued that the newly born NS accreted the surrounding matter with a very high specific angular momentum, to form an accretion disk, and a jet is likely to be launched. Combining magnetar and jets activities may take place in some peculiar superluminous CCSNe \citep{Gof19}.  

 We find that the ratio between the injected energy and the initial kinetic energy of SN ejecta $E_{\mathrm{m}}/E_{\mathrm{sn}}  \simeq 1.6-2.9$, which means that only a sufficiently powerful energy injection can make double-peaked light curves. For a powerful magnetar, when the shock expands and crosses the transition velocity point, the swept-up shell breaks up by Rayleigh–Taylor instability \citep{Blo01,Blo17} and the bubble material blows out through the shell \citep{Chen16, Suz17,Suz20}.  The escaping matter from the blowout can drive a faster shock wave into the surroundings medium, thus may produce a more prominent breakout signal \citep{Blo17}. In addition, this blowout may also hasten the radiative losses of the gas. The energy injection for the magnetar could be anisotropic with a jet-like structure \citep[e.g.,][]{Buc09}, which may lead to the breakout emission that is easier to see from the polar viewing angle. \cite{Kap2020} considered a bipolar ejecta and found that radiation can escape much easier on the polar direction, leading to a more rapid luminosity drop in the light curve. 

Because the dissipation mechanism for the magnetar wind in the medium is unclear, we make an assumption that the injected magnetar energy is thermalized spherically at the bottom of the supernova ejecta.  Due to the SN ejecta is optically thick enough for these non-thermal high-energy photons at early time, thermalization of the magnetar heating would be suppressed before $t_{\mathrm{delay}}$.  The values of $t_{\mathrm{delay}}$ which is estimated to be in range of $\simeq$ 15-43 days by comparing our model with the observational data.

In order to understand the early bump of SLSNe-I in more detail, spectra measurements at similar phases are required. Magnetar-driven shock breakout model is expected to produce a blue quasi-blackbody spectrum that is mostly featureless owing to the high temperature and ionization state.  For the extended material shock cooling model, narrow hydrogen or helium signatures are expected to be present in the early time spectra. Spectra taken during the early bump would  be helpful to diagnose different models. In addition, multi-frequency studies of double-peaked SLSNe-I are helpful in distinguishing different energy source mechanisms. For instance, \cite{Met14} suggested that the ionization breakout from the magnetar wind nebular would produce prominent X-ray emission.

 \cite{Nic16} proposed that many SLSNe-I have either been at too high redshift or did not have the cadence to detect the early bumps. However, \cite{And18} reported a  nearby type I SLSN, SN 2018bsz at 111 Mpc, which seems to lack a distinct bump but shows a long, slowly rising early plateau feature. This event suggests that the early bump may be not ubiquitous in SLSNe-I.  Determining the fraction of SLSNe-I with double-peaked light curves would help discriminate different energy source mechanisms. In this paper, we do not emphasize the completeness of samples as we used only three well-observed double-peaked SLSNe-I for the purpose of test.   Wide-area, high-cadence surveys such as the Zwicky Transient Facility (ZTF) and the Large Synoptic Survey Telescope (LSST) are thus well-suited for identifying whether or not double-peaked light curves are common in SLSNe-I. In the coming years, modeling of the double-peaked SLSNe-I based on a large sample will become more feasible.

\acknowledgments
We thank the referee for constructive suggestions.
We thank Ling-Jun Wang, Yun-Wei Yu, Chen-Han Tang, Shao-Ze Li, and Shan-Qing Wang for helpful discussion. L.D.L is supported by the National Postdoctoral Program for Innovative Talents (grant No.
BX20190044), China Postdoctoral Science Foundation
(grant No. 2019M660515) and ``LiYun” postdoctoral fellow of Beijing Normal University. H.G. is supported by the National Science Foundation of China (NSFC grants 11690024, 12021003, 11633001). X. Wang is supported by the National Science Foundation of China (NSFC grants 12033003, 11633002, and 11761141001), the Major State Basic Research Development Program (grant no. 2016YFA0400803), and the Scholar Program of Beijing Academy of Science and Technology (DZ:BS202002). S.Y. is supported by the GREAT research environment grant 2016-06012.

\bibliography{sample63}{}
\bibliographystyle{aasjournal}

\end{document}